\documentclass[twocolumn,showpacs,aps,prl,superscriptaddress]{revtex4}
\usepackage{graphicx}
\usepackage{dcolumn}
\usepackage{epsfig}
\usepackage{amsmath}
\RequirePackage{xspace}

\newcommand{\BaBarYear}    {09}
\newcommand{\BaBarNumber}  {011}
\newcommand{\SLACPubNumber} {13709}
\newcommand{\BaBarType}      {PUB}

\def\epem  {\ensuremath{e^+e^-}\xspace}

\newcommand{\costhr}{\ensuremath{\cos\thetaT}}
\newcommand{\hel}{\ensuremath{{\cal H}}}

\def\ra                 {\ensuremath{\rightarrow}\xspace}

\newcommand{\psfile}[3][]{
  \begin{center}
    \setlength{\epsfxsize}{#3\linewidth}\leavevmode
    \def\noOpt{}\def\testit{#1}\ifx\testit\noOpt%
      \epsfbox{#2}%
    \else%
      \epsfbox[#1]{#2}%
    \fi
  \end{center}
}
\def\pip   {\ensuremath{\pi^+}\xspace}
\def\pim   {\ensuremath{\pi^-}\xspace}
\def\piz   {\ensuremath{\pi^0}\xspace}
\def\Bu      {\ensuremath{B^+}}
\def\Bub     {\ensuremath{B^-}}
\def\Bbar    {\overline{B}{}}
\def\Bzb     {\ensuremath{\Bbar^0}}
\def\Bz      {\ensuremath{B^0}}
\def\BpBm    {\ensuremath{\Bu  \Bub}}
\def\BzBzb   {\ensuremath{\Bz  \Bzb}}
\def\mes{\mbox{$m_{\rm ES}$}}
\def\babar{{\em B}{\footnotesize\em A}{\em B}{\footnotesize\em AR}}
\def\pep2{PEP-II}
\def\BB      {\ensuremath{B\Bbar}\xspace}

\newcommand{\DE}{\ensuremath{\Delta E}}
\newcommand{\calB}{\mbox{${\cal B}$}}
\newcommand{\thetaT}{\ensuremath{\theta_{\rm T}}}
\newcommand{\UfourS}{\mbox{$\Upsilon(4S)$}}

\newcommand\etal{{\it et al.}}
\newcommand{\dedx}{\ensuremath{\mathrm{d}\hspace{-0.1em}E/\mathrm{d}x}}
\newcommand{\gevcc}{\mbox{$\textrm{GeV}/c^2$}} 
\newcommand{\mevcc}{\mbox{$\textrm{MeV}/c^2$}} 
\newcommand{\gevc}{\mbox{$\textrm{GeV}/c$}} 
\newcommand{\gev}{\mbox{$\textrm{GeV}$}} 
\newcommand{\mev}{\mbox{$\textrm{MeV}$}}

\newcommand{\jprlBase}  [1]     {Phys.\ Rev.\ Lett. \xspace}
\newcommand{\jprl}      [1]    {\jprlBase\  ~{\bf #1}}
\newcommand{\jprBase}        {Phys.\ Rev.\ }
\newcommand{\jprd}      [1]  {\jprBase\ D~{\bf #1}}
\newcommand{\plBase}   [1]         {Phys.\ Lett. \xspace}
\newcommand{\plb}      [1]    {\plBase\  B~{\bf #1}}

\newcommand{\npBase}         {Nucl.\ Phys.\ }
\newcommand{\npb}       [1]  {\npBase\ B~{\bf #1}}

\newcommand{\pvec}{{\bf p}}
\def\qqbar{\mbox{$q\bar q\ $}}
\newcommand{\xf}{\mbox{${\cal F}$}}

\newcommand{\aunop}{\mbox{$a_1(1260)^+  $}}
\newcommand{\aunopm}{\mbox{$a_1(1260)^{\pm}  $}}
\newcommand{\aunom}{\mbox{$a_1(1260)^-  $}}
\newcommand{\aunosemplice}{\mbox{$a_1  $}}
\newcommand{\aunopsemplice}{\mbox{$a_1^+  $}}

\newcommand{\aunomsemplice}{\mbox{$a_1^-  $}}
\newcommand{\fL}{\mbox{$f_L$}}
\newcommand{\fT}{\mbox{$f_T$}}
\newcommand{ \apam} {\ensuremath{\aunop\ \aunom\ }\xspace}
\newcommand{\RapamT}{\ensuremath{{(47.3 \pm 10.5 \pm 6.3)}}\times  10^{-6}} 
\newcommand{\RapamP}{\ensuremath{{(11.8 \pm 2.6 \pm 1.6)}}\times 10^{-6}}     
\newcommand{\Bapam}{\mbox{$B^0 \ra \aunop\ \aunom\ $}}
\newcommand{\Bapamsemplice}{\mbox{$B^0 \ra \aunopsemplice\ \aunomsemplice\ $}}
\newcommand{\BtoAA}{\mbox{$B \ra A A $}}
\newcommand{\Apipipi}{\mbox{$\aunop \ra \pim \pip \pip $}}
\newcommand{\Apipipisemplice}{\mbox{$\aunopsemplice \ra \pim \pip \pip $}}
\newcommand{\Apipipimp}{\mbox{$\aunopm \ra \pim \pip \pi^{\pm} $}}

\newcommand{\Apipizpizsemplice}{\mbox{$\aunopsemplice \ra \pip \piz \piz $}}

\newcommand{\Atrepisemplice}{\mbox{$\aunopsemplice \ra (3 \pi)^+  $}}

\newcommand{\BrBapam}{\mbox{$\calB(\Bapam )$}}
\newcommand{\BrBapamsemplice}{\mbox{$\calB(\Bapamsemplice )$}}
\newcommand{\Braunop}{\mbox{$\calB(\Apipipi )$}}

\newcommand{\Braunopsemplice}{\mbox{$\calB(\Apipipisemplice )$}}
\newcommand{\Braunozsemplice}{\mbox{$\calB(\Apipizpizsemplice )$}}
\newcommand{\BrAtrepisemplice}{\mbox{$\calB(\Atrepisemplice )$}}

\def\figurebox#1#2#3{%
    \def\arg{#3}%
    \ifx\arg\empty
    {\hfill\vbox{\hsize#2\hrule\hbox to #2{\vrule\hfill\vbox to #1{\hsize#2\vfill}\vrule}\hrule}\hfill}%
    \else
    {\hfill\epsfbox{#3}\hfill}%
    \fi}

\begin{document}

\begin{flushleft}
~\\
\end{flushleft}

\begin{flushright}
~\\
~\\
\babar-\BaBarType-\BaBarYear/\BaBarNumber \\
SLAC-\BaBarType-\SLACPubNumber \\
\end{flushright}

\title{\large  \bf\boldmath Observation and Polarization Measurement of $B^0$ \ra  \apam\ Decay}


%
\author{B.~Aubert}
\author{Y.~Karyotakis}
\author{J.~P.~Lees}
\author{V.~Poireau}
\author{E.~Prencipe}
\author{X.~Prudent}
\author{V.~Tisserand}
\affiliation{Laboratoire d'Annecy-le-Vieux de Physique des Particules (LAPP), Universit\'e de Savoie, CNRS/IN2P3,  F-74941 Annecy-Le-Vieux, France}
\author{J.~Garra~Tico}
\author{E.~Grauges}
\affiliation{Universitat de Barcelona, Facultat de Fisica, Departament ECM, E-08028 Barcelona, Spain }
\author{M.~Martinelli$^{ab}$}
\author{A.~Palano$^{ab}$ }
\author{M.~Pappagallo$^{ab}$ }
\affiliation{INFN Sezione di Bari$^{a}$; Dipartimento di Fisica, Universit\`a di Bari$^{b}$, I-70126 Bari, Italy }
% ...
\collaboration{The \babar\ Collaboration}
\noaffiliation

\begin{abstract}
We present measurements of the branching fraction \calB\ and 
longitudinal polarization fraction \fL\ for \Bapam\ decays,
with \Apipipimp. The data sample, collected with the \babar\ detector
at the SLAC National Accelerator Laboratory, represents $465 \times 10^6$
produced  \BB\ pairs.
We measure  $\BrBapam \times [\Braunop\ ]^2 = \RapamP$
and $\fL = 0.31 \pm 0.22 \pm 0.10$, where the first uncertainty is statistical
and the second systematic.
The decay mode is measured with a significance of 5.0
standard deviations including systematic uncertainties.
\end{abstract}


\maketitle

Charmless $B$ decays to  final states involving two 
axial-vector mesons (AA)  have received considerable theoretical 
attention in the last few years \cite{Cheng, Calderon}. Using QCD factorization,
the branching fractions of several  \BtoAA\ decay modes have been
calculated.
Predictions for the branching fraction of the $B^0 \ra \aunop\ \aunom$
decay mode vary between $37.4 \times 10^{-6} $ \cite{Cheng} and $6.4 \times 10^{-6} $ \cite{Calderon}.
Branching fractions at this level should be observable with the \babar\ data sample,
which can be used to discriminate between the predictions.
The predicted value of the longitudinal polarization fraction $\fL$\ is
0.64 \cite{Cheng}.
The only available experimental information on this $B$\ decay mode is the
branching fraction upper limit (UL) of $2.8 \times 10^{-3}$\ at 90\%
confidence level (CL) measured by CLEO \cite{a1a1CLEO}.

The measured value $\fL \sim 0.5$\ in penguin-dominated
$B \ra \phi K^*$ decays \cite{PhiKst}
is in contrast with naive standard model (SM) calculations
predicting a  dominant longitudinal polarization ($\fL \sim 1$)
in $B$ decays to vector-vector (VV) final  states.
The naive SM expectation is confirmed in the 
tree-dominated $B \ra \rho \rho$ \cite{RhoRho} and $B^+ \ra \omega
\rho^+$ \cite{OmegaRho} decays.  A value of 
$\fL\ \sim 1$ is found in vector-tensor $B \ra \phi
K^*_2(1430)$ decays \cite{Kst2}, while  $\fL\ \sim 0.5$
is found in $B \ra \omega K^*_2(1430)$\ decays \cite{OmegaRho}
(see Ref~\cite{PDGReview} for further discussion).

The small value of \fL\ observed in $B \ra \phi K^*$\ decays has stimulated
theoretical effort, such as the introduction of non-factorizable 
terms and penguin-annihilation amplitudes \cite{SMCal}. Other explanations 
invoke new physics \cite{NP}. Measurement of \fL\ in \Bapamsemplice~\cite{notazione}
decays will provide additional information.

We present the first measurements of the branching fraction and polarization 
in \Bapamsemplice\ decays, with \Apipipisemplice\ \cite{CC}.
We do not separate the P-wave $(\pi\pi)_{\rho}$
and the S-wave  $(\pi\pi)_{\sigma}$\ components in the 
$a_1 \rightarrow 3\pi$\ decay; a
systematic uncertainty is estimated due to the difference in the
selection efficiencies \cite{A1Pi}.
Due to the limited number of signal events expected in the
data sample, we do not perform a full angular analysis.
Using helicity formalism,  and after integration over
the azimuthal angle between the decay planes of the two  \aunosemplice\
mesons, the predicted angular distribution $d\Gamma/d\cos{\theta} $\ is:
 \begin{eqnarray}
\frac{1}{\Gamma}\frac{d\Gamma}{d\cos\theta}
\propto
\fL (1 - \cos^2{\theta}) + \frac{1}{2} \fT(1+\cos^2{\theta}),
\end{eqnarray}
where $\fT = 1 - \fL$\ and $\theta$\ is the angle between the normal to
the decay plane of
the three pions of one \aunosemplice\  and the flight direction of the other 
\aunosemplice,  both calculated in the rest frame of the first \aunosemplice.

The results presented here are based on data collected
with the \babar\ detector~\cite{BABARNIM}
at the PEP-II asymmetric-energy $e^+e^-$ collider~\cite{pep}
located at the SLAC National Accelerator Laboratory.
The analysis uses an integrated
luminosity of 423.0~fb$^{-1}$, corresponding to 
$(465 \pm 5) \times 10^6$  \BB\ pairs, recorded at the $\Upsilon (4S)$ 
resonance at a center-of-mass energy of $\sqrt{s}=10.58\ \gev$.
An additional 43.9~fb$^{-1}$, taken about 40 \mev\ below this energy (off-resonance data),
is used for the study of \qqbar\ continuum background ($\epem\ra\qqbar$, with $q = u, d, s, c$).

Charged particles are detected, and their
momenta measured, by a combination of a vertex tracker (SVT) consisting
of five layers of double-sided silicon microstrip detectors, and a
40-layer central drift chamber (DCH), both operating in the 1.5 T magnetic
field of a superconducting solenoid. The tracking system covers 92\% of
the solid angle in the center-of-mass frame. We identify photons and electrons 
using a CsI(Tl) electromagnetic calorimeter (EMC).
Further charged-particle identification is provided by the  specific energy
loss (\dedx ) in the tracking devices and by an internally reflecting
ring-imaging Cherenkov detector (DIRC) covering the central region.
A $K/\pi$ separation of better than four standard deviations 
is achieved for momenta below 3~\gevc, decreasing to 2.5~$\sigma$ at the 
highest momenta in the $B$ decay final states. A more detailed description
of the reconstruction of charged tracks in \babar\ can be found elsewhere~\cite{Aux}. 

Monte Carlo (MC) simulations of the signal decay mode,  continuum, \BB
backgrounds and detector response \cite{Geant} are used to establish
the event selection criteria. The MC signal events are simulated as \Bz\
decays to $\aunopsemplice\ \aunomsemplice$\ with $\aunosemplice \ra \rho(770) \pi$.
The \aunosemplice\ meson parameters in the simulation are:
mass $m_0 = 1230$ \mevcc\ and width $\Gamma_0 = 400$
\mevcc\ \cite{PDG2008,EvtGen}.

We reconstruct the decay of \aunopsemplice\ into three charged pions.
Two pion candidates are combined to form a $\rho^0$\ candidate.
Candidates with an invariant mass between $0.51$\ and $  1.10$ \gevcc
are combined with a third pion to form an $a_1$\ candidate.
The $a_1$\ candidate is required to have a mass between $0.87$\ and $1.75$ \gevcc.
We impose several particle identification requirements to ensure the identity of
the signal pions.
We also require the $\chi^2$ probability of the $B$\ vertex fit to be greater than
0.01 and the number of charged tracks in the event to be greater or equal to seven.

A $B$ meson candidate is kinematically characterized by the energy-substituted
mass $\mes \equiv \sqrt{(s/2 + \pvec_0\cdot \pvec_B)^2/E_0^2 - \pvec_B^2}$ and
energy difference $\DE \equiv E_B^*-\sqrt{s}/2$, where the subscripts $0$ and
$B$ refer to the initial \UfourS\ and the $B$ candidate in the laboratory frame, respectively,
and the asterisk denotes the \UfourS\ frame. The resolutions in \mes\
and \DE\ are about $3.0$~\mevcc\ and $20$ \mev, respectively.
We require candidates to satisfy $5.27 \le \mes \le 5.29$ \gevcc \ and $ -90 <  \DE  < 70$ \mev.

Background arises primarily from random track combinations in continuum events.
We reduce this background by using the angle
\thetaT\ between the thrust axis of the $B$\ candidate and the thrust axis
of the rest of the event (ROE),  evaluated in the \UfourS\ rest frame.
The distribution of $|\costhr|$ is sharply peaked near $1$ for combinations
drawn from jet-like continuum
events and is nearly uniform for \BB\ events; for this reason, we require
$|\costhr|<0.65$. 

Background can also arise from \BB\ events, especially events containing
a charmed meson (these are mostly events with five pions and a
mis-identified kaon in the final state). The charmed background includes
peaking modes, with structures in \mes\ and \DE\ that mimic signal events,
and non-peaking ``generic" modes. To suppress the charm background, we
reconstruct $D$\ and $D^*$\ mesons. Events are vetoed if they contain $D$\ or $D^*$
candidates with reconstructed masses within 20 MeV/$c^2$\ 
(window size of about $\pm$2$\sigma$) of the nominal charmed meson masses~\cite{PDG2008}.

The mean number of $B$\ candidates per event is 2.9.
If an event has multiple $B$ candidates, we select the candidate with 
the highest $B$ vertex $\chi^2$ probability. From MC simulation, we find that
this algorithm selects the correct candidate 90\% of the time
in signal events while inducing negligible bias.

Using MC simulation of signal events with longitudinal (transverse) polarization,
signal events are divided in two categories: correctly reconstructed  signal (CR), where all
candidate particles come from the correct signal \Bz , and self-cross
feed (SCF) signal, where  candidate particles are exchanged with a ROE
particle.  The fraction  of SCF candidates is $31.8\pm3.2$ ($19.4\pm1.9$)\%.

We determine the number of signal events (the signal yield) from an unbinned
extended maximum-likelihood (ML) fit. The seven input observables are \DE,
\mes , a Fisher discriminant \xf~\cite{Aux}, the two \aunosemplice\ masses and the two $\hel =|\cos{\theta}|$.
The Fisher discriminant \xf\ combines four variables calculated in the \UfourS\ frame:
the absolute values of the cosines of the angles with
respect to the beam axis of the $B$ momentum and the thrust axis of the $B$ decay products,
and the zeroth and second angular Legendre moments $L_{0,2}$ 
of the momentum flow about the $B$ thrust axis. The Legendre moments are defined by
$ L_k = \sum_m p_m  \left|\cos\theta_m\right|^k$,
where $\theta_m$ is the angle with respect to the $B$ thrust axis of a
track or neutral cluster $m$, $p_m$ is its momentum, and the sum
includes the ROE particles only.

There are five hypotheses in the likelihood model: signal, continuum,
and three  \BB\ components, which take into account
charmless, generic charm and peaking charm backgrounds.
The likelihood function is:
\begin{equation}
{\cal L}=  e^{-\left(\sum_{j=1}^5 n_j\right)} \prod_{i=1}^N \left[\sum_{j=1}^5 n_j  {\cal P}_j ({\bf x}_i)\right],
\end{equation}
where $N$ is the number of input events,  $n_j$ is the number of events for 
hypothesis $j$ and     ${\cal P}_j ({\bf x}_i)$ is the corresponding
probability density function (PDF), evaluated   
with the observables ${\bf x}_i$ of the $i$th event. 
Since   correlations among the observables are small ($<10$\%),
we take each  ${\cal P}$  as the product of the PDFs for the separate
variables.

The signal includes both CR and SCF signal  components with 
the SCF fraction fixed in the fit to the value estimated from MC
 simulation. Both CR and SCF signals are used to measure the branching
 fraction and polarization. The PDF of the signal takes the form:
\begin{eqnarray}
P_{sig} & = &  \fL \left(1- g_L^{SCF}\right) {\cal P}_{CR,L} + \fL g_L^{SCF} {\cal P}_{SCF,L} \\
        & + &  \fT \left(1- g_T^{SCF}\right) {\cal P}_{CR,T} + \fT g_T^{SCF} {\cal P}_{SCF,T} \, \nonumber
\end{eqnarray}
where  $g_L^{SCF}$\ ($g_T^{SCF}$) is the fraction of SCF in longitudinal (transverse) polarized signal events
and   ${\cal P}_{CR,L}$,  ${\cal P}_{SCF,L}$ 
(${\cal P}_{CR,T}$,  ${\cal P}_{SCF,T}$) are the signal PDFs of CR and SCF signal 
components for  longitudinal (transverse) polarization.

We determine the PDF parameters from Monte Carlo simulation for the
signal and \BB\ backgrounds and from off-resonance data for the
continuum background.

We parameterize \mes\ and \DE\ using a Gaussian function with exponential
tails~\cite{cruj} for the CR signal and charmless components,
and using polynomials for all other components, except for the \mes\ distribution for continuum events
which is described by the ARGUS empirical phase space function~\cite{argus} $x\sqrt{1-x^2}\exp{\left[-\xi(1-x^2)\right]}$, where $x\equiv2\mes/\sqrt{s}$\ and $\xi$\ is a parameter.
The \aunosemplice\ mass is described by a relativistic Breit-Wigner
function for the CR signal component, an asymmetric Gaussian plus a linear
polynomial for the SCF signal component, and polynomials for the remaining
components. The Fisher variable is parametrized with an
asymmetric Gaussian plus a linear polynomial in all cases. The $\cal H$
variables are parametrized  with a Gaussian plus a linear polynomial for
the charm peaking component and with a polynomial in all other cases.
The parameters left free in the fit are the signal, continuum,
and three \BB\ component yields, and \fL. We also float some of the parameters
of the continuum PDFs:
the three parameters of the asymmetric Gaussian part of \xf,
and one parameter each for the \hel, the $a_1$\ masses and \DE. 

Large data samples of $B$ decays to charmed final states ($B^0 \rightarrow {D^*}^- a_1^+$),
which have similar topology to the signal, are used to verify the simulated resolutions in \mes\ and \DE.
Where the data samples reveal differences from the Monte Carlo
we shift or scale the resolution function used in the likelihood fits.
Any bias in the fit, which arises mainly from neglecting the small
correlations among the discriminating observables, is determined from a large set of simulated
experiments for which the continuum background is generated from the PDFs, and
into which we have embedded  
the expected number of \BB\  background, signal and SCF  events 
chosen  randomly from fully simulated Monte Carlo samples.

\begin{table}[!htp]
\caption{
Fitted signal yield and  yield bias (in events), bias on \fL, detection
efficiencies   $\epsilon_L $  and
$\epsilon_T $  for events with longitudinal and
transversal polarization, respectively, significance $S$ 
(including systematic uncertainties), measured branching
fraction \calB\ and fraction of longitudinal polarization \fL\ with
statistical and systematic uncertainties.}
\label{tab:results}
\begin{center}
\begin{tabular}{lc}
\hline\hline
Signal yield & $545 \pm 118$ \\
Signal yield bias & $ +14$ \\
\fL bias & $ -0.06$\\
$\epsilon_L$ (\%) &$9.0$\\
$\epsilon_T$ (\%) &$10.0$\\
$S$ $(\sigma)$ & $5.0$ \\
$\calB\ (\times 10^{-6})$ & $11.8 \pm 2.6 \pm 1.6$ \\
\fL & $0.31 \pm 0.22 \pm 0.10 $\\
\hline\hline 
\end{tabular}
\end{center}
\end{table}

The fit results are presented in Table~\ref{tab:results}.
The detection efficiencies  are 
calculated as the ratio of the number of signal MC events
passing all the cuts to the total number generated.
We compute the branching fraction by subtracting the fit bias from the
measured yield, and dividing the result by the number of produced \BB\
pairs times the product of the daughter branching fractions and  the 
detection efficiency. We assume that the
branching fractions of the \UfourS\ to \BpBm\ and \BzBzb\ are each 50\%.
The branching fraction and \fL\
are corrected for the slightly different reconstruction efficiencies in
longitudinal and transversal polarizations.
The statistical uncertainty on the signal yield is taken as the change in 
the central value when the quantity $-2\ln{\cal L}$ increases by one 
unit from its minimum value.
The significance is  the square root 
of the difference between the value of $-2\ln{\cal L}$ (with systematic 
uncertainties included) for zero signal and the value at its
minimum.
In this calculation we have taken into account the fact that the 
floating $\fL$\ parameter is not defined in the zero
signal hypothesis.

\begin{figure}[t]
\resizebox{\columnwidth}{!}{
\begin{tabular}{cc}
\includegraphics[]{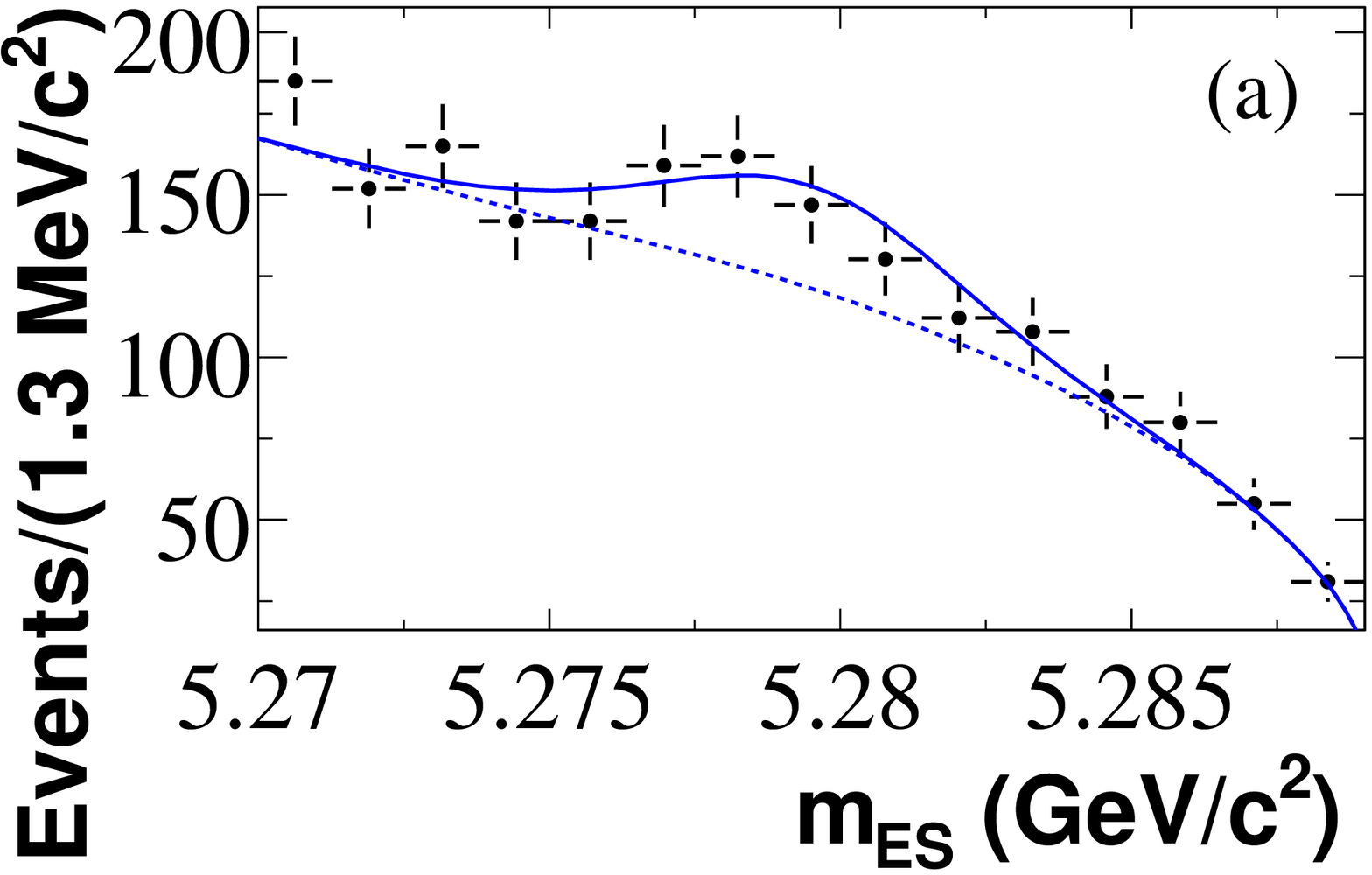} &
\includegraphics[]{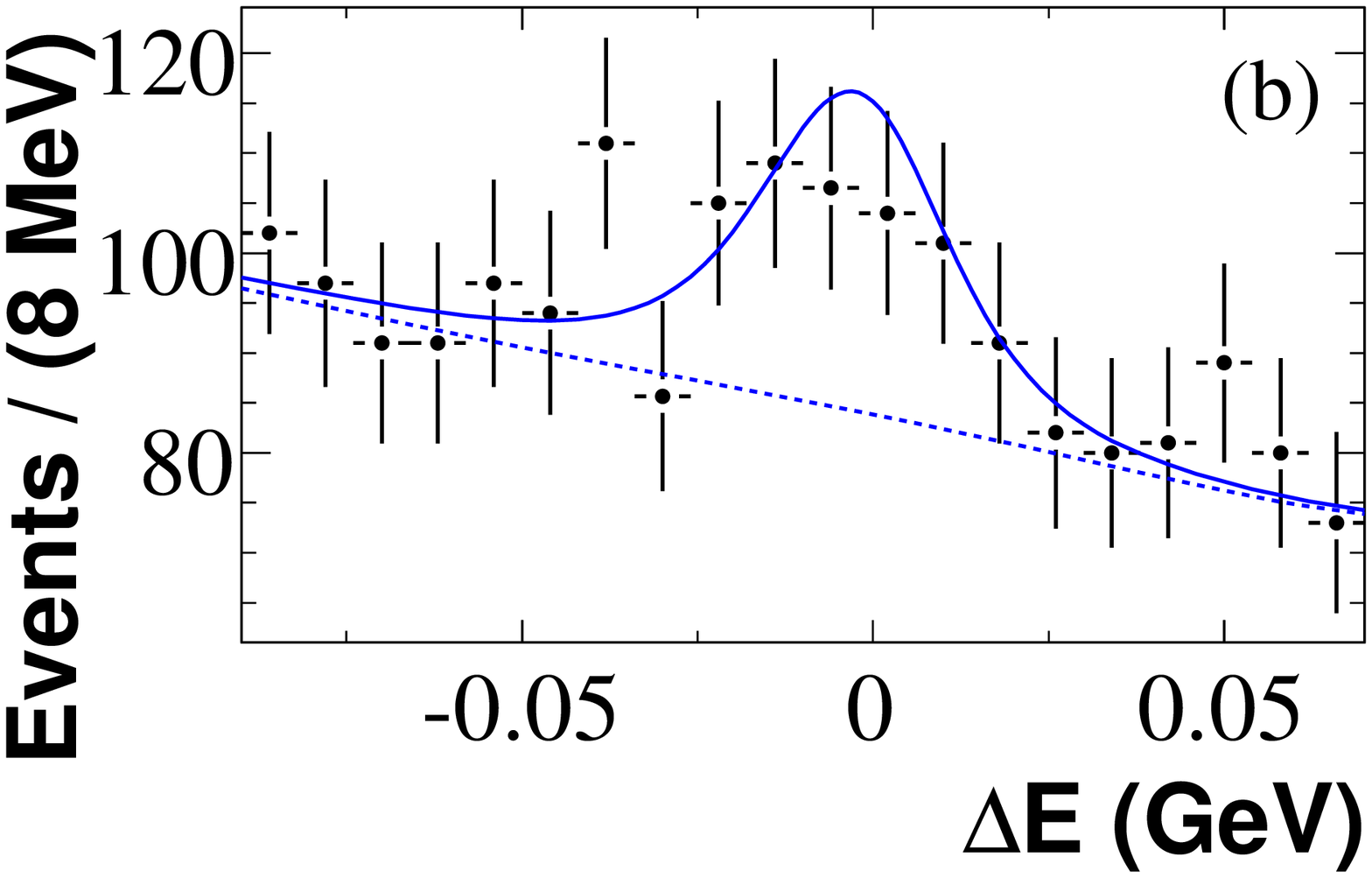} \\
\includegraphics[]{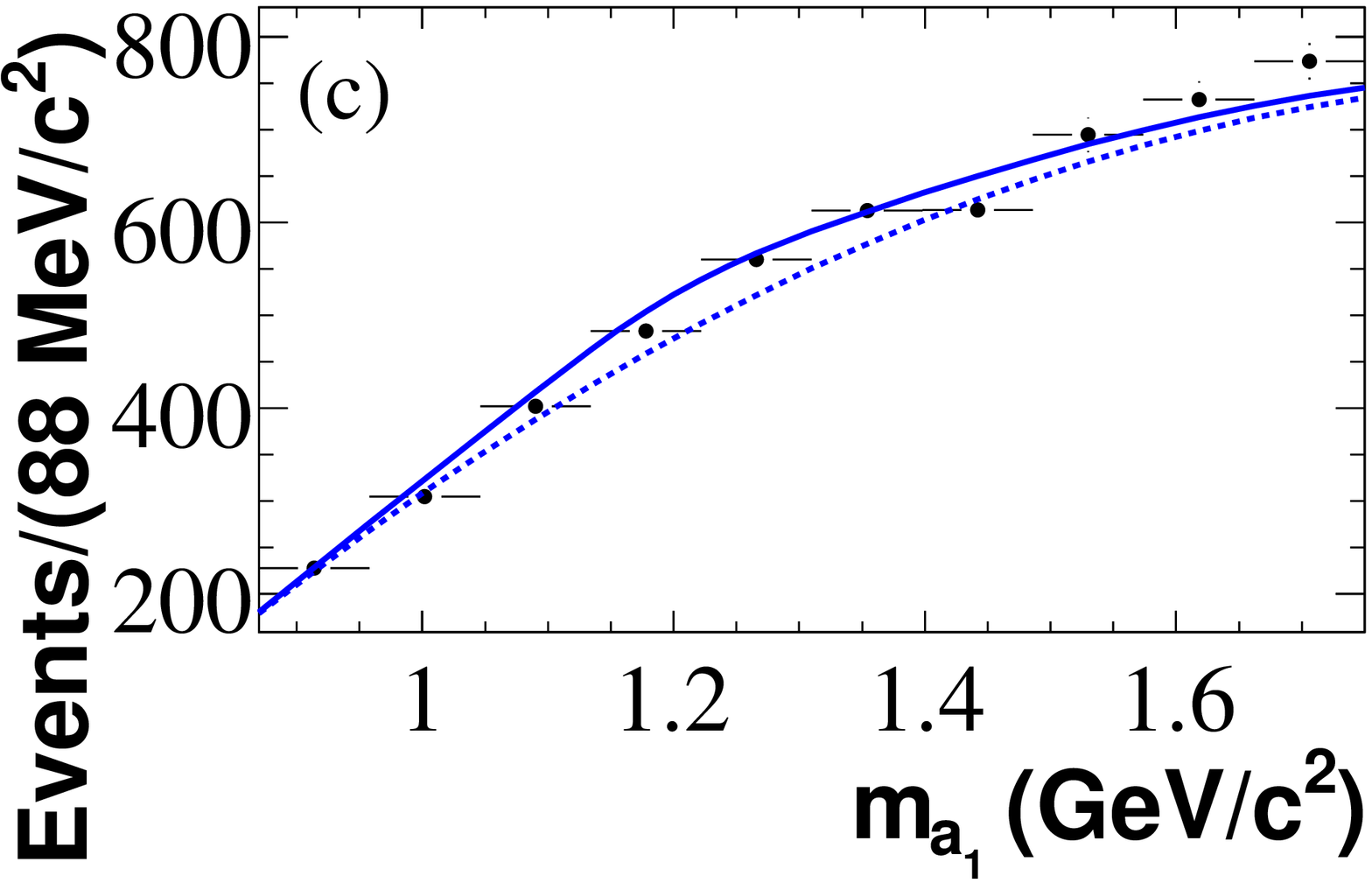} &
\includegraphics[]{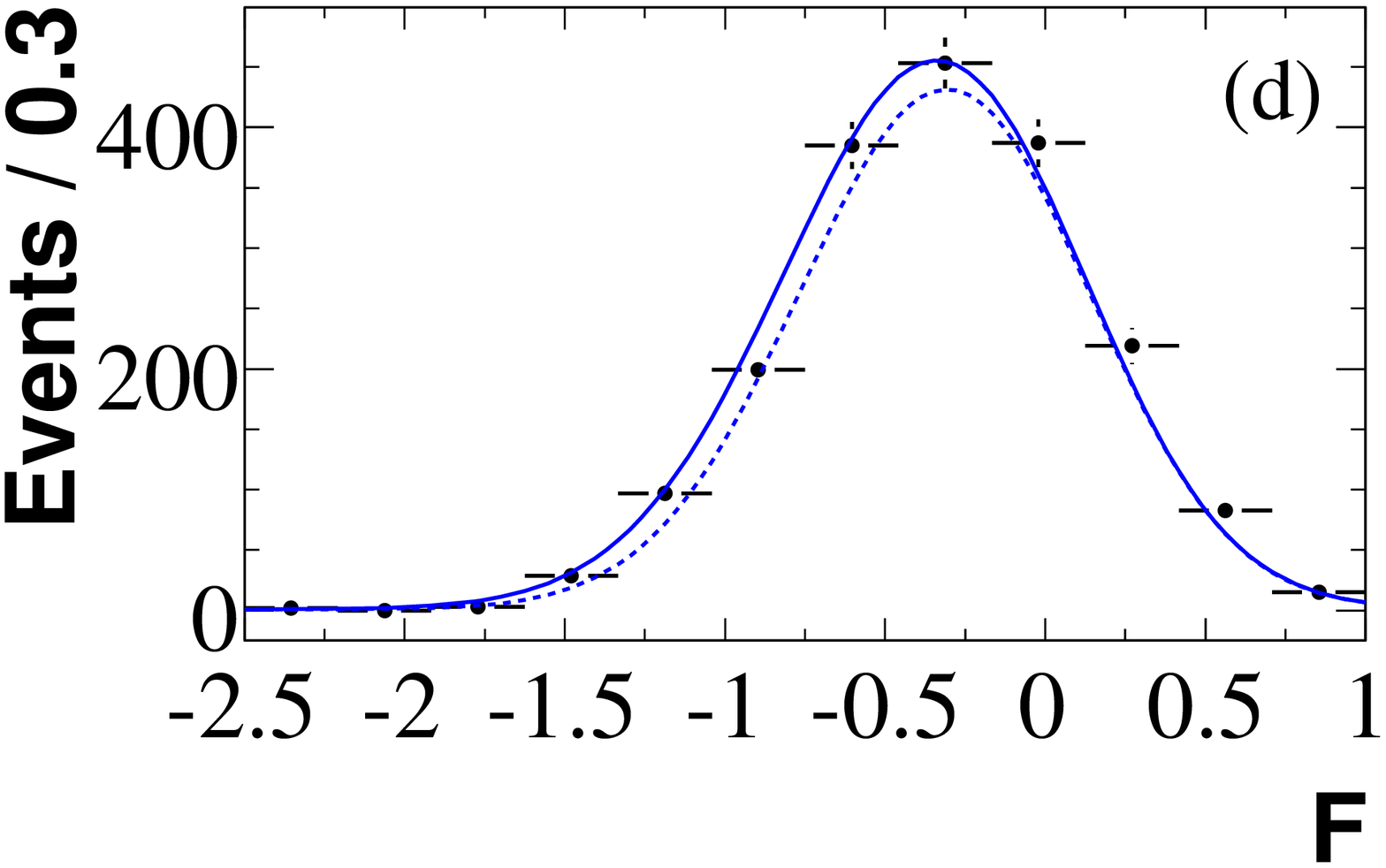} \\
\multicolumn{2}{c}{\includegraphics[]{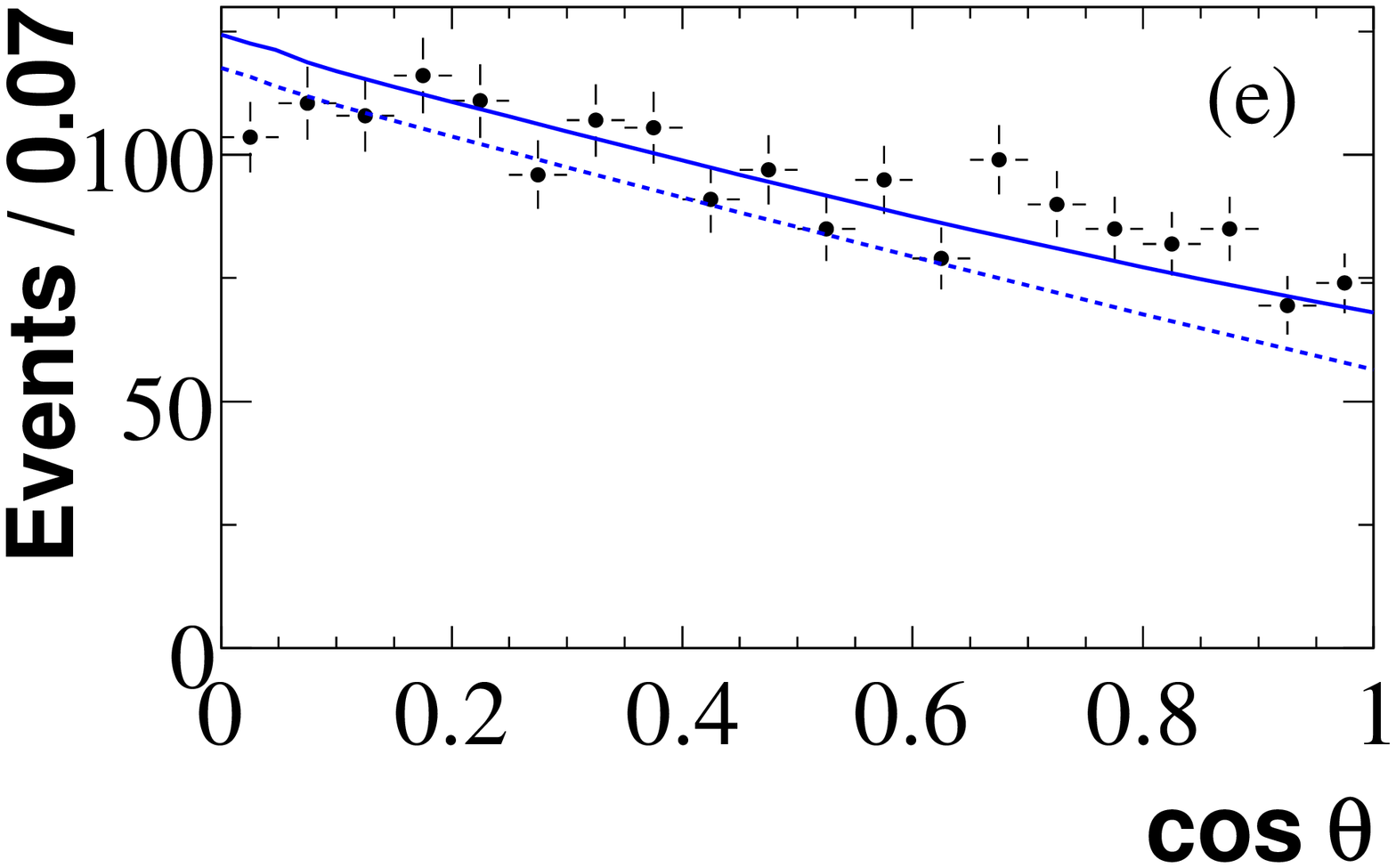}}
\end{tabular}
}
\caption{\label{fig:projections}
Projections of (a) \mes, (b) \DE, (c) \aunosemplice\ invariant mass (average of
$m_{a_1^+}$\ and $m_{a_1^-}$\ is shown), (d) \xf\ and (e) $\hel =|\cos{\theta}|$ 
(average of $|\cos{\theta}_{a_1^+}|$\ and $|\cos{\theta}_{a_1^-}|$\ is shown).
Points with error bars (statistical only) represent the data, the solid
line the full fit function, and the dashed line the background component.
These plots are made with a requirement on the signal likelihood that selects
25\%-40\% of the signal and 2\%-5\% of the background.
}
\label{fig:sPlots}
\end{figure}

Figure~\ref{fig:projections} shows 
the  projections of \mes\ , \DE\ , the \aunosemplice\ invariant mass,
\xf\ and  $\hel$
for a subset of the data for which the ratio of the signal likelihood to the total 
likelihood (computed without using the variable plotted) exceeds a 
threshold that optimizes the sensitivity.


A systematic uncertainty of 38 events on the signal yield
 due to  the PDF  parametrization  is estimated by varying the 
signal PDF parameters within their  uncertainties,  obtained through comparison of MC
and data in control samples.
The uncertainty from the fit bias (7 events) is taken as half the correction
itself.
Uncertainty from lack of  knowledge of the \aunosemplice\ meson parameters
is  31 events.
We vary the SCF fractions by their uncertainties
and estimate a systematic uncertainty of 12 events.
A systematic uncertainty of 19 events
from possible contamination by $B^0 \ra \aunop\ a_2(1320)^-$\ background
events is estimated with simulated MC experiments.
The uncertainty due to cross feed between the signal and non-resonant
backgrounds, evaluated with MC events, is 10 events.
Uncertainties of 1.4\% and 3.6\%  are associated with the track efficiency
and particle identification, respectively.
Differences between data and simulation for the \costhr\ variable
lead to a systematic uncertainty of 2.5\%.
Assuming that 20\% of \aunosemplice\ decays proceed through
 the S-wave $(\pi\pi)_\sigma$\ channel~\cite{PDG2008},
we estimate a systematic uncertainty of 
6.8\%  from the difference in
reconstruction efficiency between the P-wave $(\pi \pi)_{\rho}$ and S-wave components.
The uncertainty in the total number of \BB\ pairs in the
data sample is 1.1\%.
The total systematic uncertainty, obtained by adding the individual terms
in quadrature, is 12.9\%.

The main systematic uncertainties on \fL\  arise from the fit bias ($0.03$), the
variation of PDF parameters ($0.08$), the \aunosemplice\ parametrization ($0.04$) 
and the non-resonant background (0.02).

In conclusion, we have measured the branching fraction:
 $\BrBapamsemplice \times [\calB ( \Atrepisemplice )]^2 = \RapamP$
and the fraction of longitudinal polarization $\fL = 0.31 \pm 0.22 \pm
 0.10$.  Assuming that 
\Braunopsemplice\ is equal to  \Braunozsemplice, and that \BrAtrepisemplice\ is equal to 100\%
\cite{PDG2008}, we obtain  $\BrBapamsemplice = \RapamT$.
The decay mode is seen with a significance of 5.0~$\sigma$\ including systematic
uncertainties.
The measured branching fraction and longitudinal polarization are
in general agreement with the theoretical expectations in~\cite{Cheng},
while they disfavor those in~\cite{Calderon}.

We are grateful for the excellent luminosity and machine conditions
provided by our \pep2\ colleagues, 
and for the substantial dedicated effort from
the computing organizations that support \babar.
The collaborating institutions wish to thank 
SLAC for its support and kind hospitality. 
This work is supported by
DOE
and NSF (USA),
NSERC (Canada),
CEA and
CNRS-IN2P3
(France),
BMBF and DFG
(Germany),
INFN (Italy),
FOM (The Netherlands),
NFR (Norway),
MES (Russia),
MEC (Spain), and
STFC (United Kingdom). 
Individuals have received support from the
Marie Curie EIF (European Union) and
the A.~P.~Sloan Foundation.


\begin{thebibliography}{99}

\bibitem{Cheng}
H.-Y. Cheng and K.-C. Yang, \jprd{78}, 094001 (2008).

\bibitem{Calderon}
G. Calderon  \etal, \jprd{76}, 094019 (2007).  

\bibitem{a1a1CLEO}
CLEO Collaboration, D. Bortoletto \etal,  \jprl{62}, 2436 (1989).

\bibitem{PhiKst}
\babar\ Collaboration, B. Aubert \etal,   \jprd{78}, 092008 (2008);
Belle Collaboration,  K. F. Chen \etal,   \jprl{91}, 201801 (2003).

\bibitem{RhoRho}
Belle Collaboration,   A. Somov \etal,  \jprl{96}, 171801 (2006);
\babar\ Collaboration, B. Aubert \etal,  \jprd{76}, 052007 (2007);
Belle Collaboration,   J. Zhang \etal,  \jprl{91}, 221801 (2003);
\babar\ Collaboration, B. Aubert \etal,  arXiv:0901.3522[hep-ex],
submitted to \jprl{}

\bibitem{OmegaRho}
\babar\ Collaboration, B. Aubert \etal,  \jprd{79}, 052005.

\bibitem{Kst2}
\babar\ Collaboration, B. Aubert \etal,  \jprl{98}, 051801 (2007).

\bibitem{PDGReview}
A. V.  Gritsan and J. G. Smith, ``Polarization in $B$  Decays''
review in  \cite{PDG2008}, \plb{667}, 910 (2008).


\bibitem{SMCal}
A. L. Kagan, \plb{\textbf{601}}, 151 (2004);
C. W. Bauer \etal, \jprd{70}, 054015 (2004);
P. Colangelo, F. De Fazio, and T. N. Pham, \plb{597}, 291 (2004);
M. Ladisa \etal, \jprd{70}, 114025 (2004);
H. Y. Cheng, C. K. Chua, and A. Soni, \jprd{71}, 014030 (2005);
H. N. Li and S. Mishima, \jprd{71}, 054025 (2005);
C. H. Chen \etal, \jprd{72}, 054011 (2005);
M. Beneke \etal, \npb{774}, 64 (2007).


\bibitem{NP}
A. K. Giri and R. Mohanta, \jprd{69}, 014008 (2004);
E. Alvarez \etal, \jprd{70}, 115014 (2004);
P. K. Das and K. C. Yang, \jprd{71}, 094002 (2005);
C. H. Chen and C. Q. Geng, \jprd{71}, 115004 (2005);
Y. D. Yang, R. M. Wang, and G. R. Lu, \jprd{72}, 015009 (2005); 
C. S. Hunger \etal, \jprd{73}, 034026 (2006);
C. H. Chen and C. Q. Geng, \jprd{75}, 054010 (2007). 

\bibitem{notazione}
$a_1$\ will be used to indicate the $a_1(1260)$\ meson.

\bibitem{CC}
Charge conjugate decay modes are implied unless specifically stated.

\bibitem{A1Pi}
\babar\ Collaboration, B. Aubert \etal, \jprl{97}, 051802 (2006).


\bibitem{BABARNIM}
\babar\ Collaboration, B.\ Aubert \etal,  Nucl. Instrum. Methods Phys. Res., Sect. A \textbf{479}, 1 (2002).

\bibitem{pep}
PEP-II Conceptual Design Report, SLAC-R-418 (1993).

\bibitem{Aux}
\babar\ Collaboration, B.~Aubert \etal,    \jprd{70}, 032006 (2004).

\bibitem{Geant}
The \babar\ detector MC simulation is based on {\sc Geant 4}: S.~Agostinelli
\etal, Nucl. Instrum. Methods Phys. Res., Sect. A \textbf{506}, 250 (2003).

\bibitem{PDG2008}
Particle Data Group, C. Amsler \etal,  \plb{667}, 1  (2008).

\bibitem{EvtGen}
EvtGen particle decay simulation package, D.~J.~Lange, Nucl. Instrum. Methods Phys. Res., Sect. A \textbf{462}, 152 (2001).

\bibitem{cruj}
We use the following function:
$$f(x)=\exp \left(
\frac{-(x-\mu)^2}{2\sigma^2_{L,R} + \alpha_{L,R} (x-\mu)^2} \right)$$
where $\mu$ is the peak position of the distribution, $\sigma_{L,R}$
are the left and right widths, and $\alpha_{L,R}$ are the left and
right tail parameters.

\bibitem{argus}
ARGUS Collaboration, H.\ Albrecht \etal, \plb{241}, 278 (1990).


\end{thebibliography}
\end{document}